\documentclass[conference]{IEEEtran}

\usepackage{setspace}

\usepackage{graphicx}
\usepackage{calc}
\usepackage{multicol}
\usepackage{apalike}
\usepackage{url}
\usepackage[ruled,linesnumbered]{algorithm2e}
\usepackage{color}
\usepackage{caption}

\begin{document}


\hyphenation{HarmonicIO} 



\title{Apache Spark Streaming, Kafka and HarmonicIO: A~Performance Benchmark and Architecture Comparison for Enterprise and Scientific Computing}

\author{
\IEEEauthorblockN{
Ben Blamey\IEEEauthorrefmark{1}, Andreas Hellander\IEEEauthorrefmark{1} and Salman Toor\IEEEauthorrefmark{1}
}
\IEEEauthorblockA{
\IEEEauthorrefmark{1} Department of Information Technology, Division of Scientific Computing, Uppsala University, Sweden\\ 
Email: \{Ben.Blamey, Andreas.Hellander, Salman.Toor\}@it.uu.se 
}
}

\maketitle

\begin{abstract}

This paper presents a benchmark of stream processing throughput comparing Apache Spark Streaming (under file-, TCP socket- and Kafka-based stream integration), with a prototype P2P stream processing framework, \emph{HarmonicIO}.
Maximum throughput for a spectrum of stream processing loads are measured, specifically, those with large message sizes (up to 10MB), and heavy CPU loads -- more typical of scientific computing use cases (such as microscopy), than enterprise contexts.
A detailed exploration of the performance characteristics with these streaming sources, under varying loads, reveals an interplay of performance trade-offs, uncovering the boundaries of good performance for each framework and streaming source integration. We compare with theoretic bounds in each case.
Based on these results, we suggest which frameworks and streaming sources are likely to offer good performance for a given load.
Broadly, the advantages of Spark's rich feature set comes at a cost of sensitivity to message size in particular -- common stream source integrations can perform poorly in the 1MB-10MB range.
The simplicity of HarmonicIO offers more robust performance in this region, especially for raw CPU utilization.




\end{abstract}




\section{Introduction}\label{intro}




A number of stream processing frameworks have gained wide adoption over the last decade or so (Apache Flink~\cite{carboneApacheFlinkStream2015}, Apache Spark Streaming~\cite{zahariaApacheSparkUnified2016}, Flume~\cite{apacheflumeApacheFlume2016}); suitable for high-volume, high-reliability stream processing workloads. Their development has been motivated by analysis of data from cloud, web and mobile applications. For example, Apache Flume is designed for the analysis of server application log data. Apache Spark improves upon the Apache Hadoop framework~\cite{apachesoftwarefoundationApacheHadoop2011} for distributed computing, and was later extended with streaming support. Apache Flink was later developed primarily for stream processing.


These frameworks boast 
performance, scalability, data security, processing guarantees, and efficient, parallelized computation; together with high-level stream processing APIs (augmenting familiar map/reduce with stream-specific functionality such as windowing). All of which makes them attractive for scientific computing -- including imaging applications in the life-sciences, and simulation output more generally.

Previous studies have shown that these frameworks are capable of processing message streams on the order of 1 million or more messages per second, but focus on enterprise use cases with textual rather than binary content, and of message size perhaps a few KB. Additionally, the computational cost of processing an individual message may be relatively small (e.g. parsing JSON, and applying some business logic). By contrast, in scientific computing domains messages can be much larger (order of several MB).



Our motivating use case 
is the development of a cloud pipeline for the processing of streams of microscopy images, for biomedical research applications. Existing systems for working with such datasets have largely focused on offline processing: our online processing (processing the `live' stream), 
is relatively novel for the image microscopy domain. 
Electron microscopes generate high-frequency streams of large, high-resolution image files (message sizes 1-10Mb), and feature extraction is computationally intensive. This is typical of many scientific computing use cases: where files have binary content, 
with execution time dominated by the per-message `map' stage.
Thereby, we investigate how well the performance of enterprise-grade stream processing frameworks (such as Apache Spark) translates to loads more characteristic of scientific computing, for example, microscopy image stream processing, by benchmarking under a spectrum of conditions representative of both. We do this by varying both the processing cost of the map stage, and the message size to expand on previous studies.


For comparison, we measured the performance of HarmonicIO~\cite{torruangwatthanaHarmonicIOScalableData2018a} -- a research prototype with a which has a P2P-based architecture, under the same conditions. 
This paper contributes:

\begin{itemize}
\item A performance comparison of an enterprise grade framework (Apache Spark) for stream processing to a streaming framework tailored for scientific use cases (HarmonicIO).

\item An analysis of these results, and comparison with theoretical bounds -- relating the findings to the architectures of the framework when integrated with various streaming sources. We find that performance varies considerably according to the application loads -- quantifying where these transitions occur.

\item Benchmarking tools for Apache Spark Streaming, 
with tunable message size and CPU load per message -- to explore this domain as a continuum.

\item Recommendations for choosing frameworks and their integration with stream sources, especially for atypical stream processing applications, highlighting some 
 limitations of both frameworks especially for scientific computing use cases.  
 
\end{itemize}



\section{BACKGROUND: STREAM PROCESSING OF IMAGES IN THE HASTE PROJECT}\label{background}


High-throughput ~\cite{wollmanHighThroughputMicroscopy2007}, and high-content imaging (HCI) experiments are highly automated experimental setups which are used to screen molecular libraries and assess the effects of compounds on cells using microscopy imaging. Work on smart cloud systems for prioritizing and organizing data from these experiments is our motivation for considering streaming applications where messages are relatively large binary objects (BLOBs) and where each processing task can be quite CPU intensive.


Our goal of \emph{online} analysis of the microscopy image stream allows both the quality of the images to analyzed (highlighting any issues with the equipment, sample preparation, etc.) as well as detection of characteristics (and changes) in the sample itself \emph{during the experiment}. 
Industry setups for high-content imaging can produce 38 frames/second with image sizes on the order of 10Mb \cite{lugnegaard2018building}. 
These image streams, like other scientific use cases, have different characteristics than many enterprise stream analytics applications: 


\begin{itemize}
\item Messages are binary (not textual, JSON, XML, etc.).
\item Messages are larger (order MBs, not bytes or KBs).
\item The initial map phase can be computationally expensive, and perhaps dominate execution time.
\end{itemize}

Our goal is to create a general pipeline able to process streams with these characteristics (and image streams in particular) with an emphasis on spatial-temporal analysis. Our SCaaS -- \emph{Scientific Computing as a Service} platform will to allow domain scientists to work with large datasets in an economically efficient way
without needing to manage infrastructure and software themselves. Apache Spark Streaming (ASS) has many of the features needed to build such a platform, with rich APIs suitable for scientific applications, and proven performance for small messages with computationally light map tasks. However, it is not clear how well this performance translates to the regimes of interest to the HASTE project. This paper explores the performance of ASS for a wide range of application characteristics, and compares it to a research prototype streaming framework HarmonicIO. 



\section{EXISTING BENCHMARKING STUDIES}\label{benchmarking}
Several studies have investigated the performance of Spark, Flink and related platforms. However, these studies tend use small messages, with a focus on sorting, joining and other stream operations.  Under an experimental setup modeling a typical enterprise stream processing pipeline \cite{chintapalliBenchmarkingStreamingComputation2016}, Flink and Storm were found to have considerably lower latencies than Spark (owing to its micro-batching implementation), whilst Spark's throughput was significantly larger. The input was small JSON documents for which the initial map -- i.e. parsing -- is cheap, and integrated the stream processing frameworks under test with Kafka~\cite{kreps2011kafka} and Redis~\cite{salvatoresanfilippoRedis2009} -- advantageous in that it models a realistic enterprise system, but with each component having its own performance characteristics, it makes it difficult to get a sense of maximum performance of the streaming frameworks in isolation. In their study the data is preloaded into Kafka. In our study, we investigate ingress bottlenecks by writing and reading data through Kafka \emph{during the benchmarking}, to get a full measurement of sustained throughput.

In an extension of their study \cite{grierExtendingYahooStreaming2016}, Spark was shown to outperform Flink in terms of throughput by a factor of 5, achieving frequencies of more than 60MHz. Again, as with previous studies, Kafka integration was used, with small messages.
Other studies follow a similar vein: \cite{qianBenchmarkingModernDistributed2016} used small messages (60 bytes, 200 bytes), and lightweight pre-processing (i.e. `map') operations: e.g. grepping and tokenizing strings, with an emphasis on common stream operations such as sorting. 
Indeed, sorting is seen as something of a canonical benchmark for distributed stream processing. For example, Spark previously won the GraySort contest \cite{xinApacheSparkFastest2014}, where the frameworks ability to shuffle
data between worker nodes is exercised.
Marcu et. al. (2016)\nocite{marcuSparkFlinkUnderstanding2016} offer a comparison of Flink and Spark on familiar BigData benchmarks (grepping, wordcount
), and give a good overview of performance optimizations in both frameworks. 

HarmonicIO, a research prototype streaming framework with a peer-to-peer architecture, developed 
specifically for scientific computing workloads, has previously shown good performance messages in the 1-10MB range \cite{torruangwatthanaHarmonicIOScalableData2018a}. To the authors' knowledge there is no existing work benchmarking stream processing with Apache Spark, or related frameworks, with messages larger than a few KB, and with map stages which are computationally expensive.


\section{STREAM PROCESSING FRAMEWORKS}\label{frameworks}

This section introduces the two frameworks selected for study in this paper, Apache Spark and HarmonicIO.
Apache Spark competes with other frameworks such as Flink, Flume, Heron in offering high performance at scale, with features relating to data integrity (such as tunable replication), processing guarantees, fault tolerance, checkpointing, and so on. HarmonicIO is a research prototype -- much simpler in implementation, and is built around direct use of TCP sockets for direct, high-throughput P2P communication.

\subsection{Apache Spark}


Apache Spark stands out as an attractive framework for scientific computing applications with its high-level APIs, such as built-in support for scalable machine learning. 
Originally developed for batch operations, in-memory caching of intermediate results improves on the performance of Hadoop, where data is typically written to a distributed file system at each stage.
Spark allows a more interactive user experience, with ad-hoc analysis, which was difficult with Hadoop. This smart caching behavior is combined with functionality to track the lineage of the calculations, allowing deterministic re-calculation in the event of errors and node failure, together with a host of other features, built around the Resilient Distributed Dataset (RDD)~\cite{Zaharia:2012:RDD:2228298.2228301}. 
Spark can scale successfully to 1000s of nodes~\cite{xinApacheSparkFastest2014}.

Spark Streaming was a later addition, leveraging the batch functionality for a streaming context by creating a new batch every few seconds (the batch interval). As with batch operations, data is further partitioned for distribution and scheduling. The Streaming API augments map/reduce operators (familiar from batch processing) with stream functionality, such as windowing. 

\subsection{HarmonicIO}

HarmonicIO~\cite{torruangwatthanaHarmonicIOScalableData2018a} is a peer-to-peer distributed processing framework, intended for high throughput of medium and large messages. 
HarmonicIO's smart architecture will favor P2P message transfer, but fall back to a queue buffer when necessary to absorb fluctuations in input or processing rate. Messages are processed in a simple loop: pop a message from the master queue (if any exists) otherwise wait to receive a message directly from the streaming source over TCP; process it; and repeat. Per-node daemons aggregate availability information at the master, from where clients query the status of available processing engines. The master node manages the message buffer queue.
Being a research prototype, it lacks many features of more mature frameworks such as resilience, error handling, guaranteed delivery etc., and does not yet offer \emph{reduce} functionality. Yet the simplicity of the implementation makes it easily adoptable and extensible.




\section{THEORETICAL BOUNDS ON PERFORMANCE}\label{theobounds}

\begin{figure}[h]
\begin{center}
\includegraphics[width=0.5\linewidth]{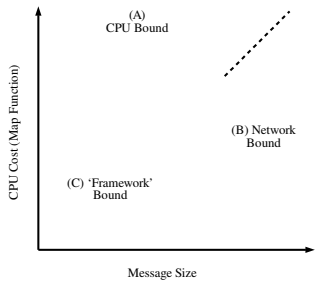}
\end{center}
\caption{Schematic of parameter space for this study, showing the processing cost of the map function, and the message size.}
\label{fig:prediction}
\end{figure}

For a range of message sizes and CPU costs of the map function,
we consider the theoretical performance of an `ideal' stream processing framework which exhibits performance equal to the tightest bound, either network or CPU, with zero overhead; across this domain. 
In this article we investigate how close the frameworks under study can approach these bounds over our domain.  
Fig.~\ref{fig:prediction} illustrates how we might expect an ideal framework to perform in different regimes.  

\textbf{A - Small message size, large processing cost - CPU Bound}: For sufficiently large processing cost in relation to message size, performance will be CPU bound. Relative performance of the frameworks in this region will be determined by their ability to utilize CPU, and minimizing processing overheads. This regime would be typical of scientific computing applications involving e.g. a simulation step as part of the map stage.

\textbf{B - Large message size, small processing cost - Network Bound}: For sufficiently large message size, performance will be network bound. In this region, relative performance between frameworks will be determined by the network topology. P2P network topologies should perform well, whereas routing messages among the worker nodes will create additional network traffic which could impair performance. This regime would be typical for scientific computing applications involving relatively simple filtering operations on binary large objects (BLOBs), such as filtering of massive genomics datasets~\cite{ausmeesBAMSIMulticloudService2018}.


\textbf{C - Small messages, small processing cost}: In this regime processing frequency should be high. This region will expose any limitations on the absolute maximum message frequency for the particular integration and thus may be `framework bound'. Well-performing frameworks may be able to approach the network bounds; with very high frequencies. This regime would be typical for the type of enterprise applications studied in previous benchmarks.

\section{METHODOLOGY}\label{method}


Varying message size (100 bytes -- 10Mb), and CPU cost for processing each message (0 -- 1 second per message)
allowed us to sample the performance of the studied streaming frameworks over a parameter space with settings ranging from highly CPU-intensive workloads to highly data-movement intensive use-cases. This domain captures typical enterprise use cases, and scientific computing workloads.




The microscopy use case is a particular focus: message (image) sizes 1-10Mb, with a CPU cost profiled at around 100mS for very simple analysis (consistent with \cite{torruangwatthanaHarmonicIOScalableData2018a}) -- CPU cost would depend on the specific features being extracted, and could be significantly higher.
We measure maximum sustained frequency (i.e throughput) at each point (message size, CPU), for each of the framework stream source integrations explained below.





\subsection{Apache Spark Streaming Source Integrations}

There are many possible stream integrations for Apache Spark Streaming -- we discuss some of the most common, likely to be adopted by new users:




\textbf{Spark + TCP Socket:} TCP sockets are a simple, universal mechanism, easy to integrate into existing applications, with minimal configuration. 


\textbf{Spark + Kakfa:} Kafka is a stream processing framework in its own right: Kafka producers write messages onto one end of message queues, which are processed by Kafka \emph{consumers}. 
Kafka is commonly used in conjunction with Spark Streaming in enterprise contexts, to provide a resilient and durable buffer, allowing Spark applications to be restarted without interrupting the streaming source.  
The newer \emph{Direct DStream} integration approach with Kafka is used in this study. 

\textbf{Spark + File Streaming:} Under this approach, Spark will process new files in each batch interval. 
We configure an NFS share on the streaming source node. This allows direct transfer of file contents from the streaming source node, to the processing machines -- similar to the TCP socket approach used in HarmonicIO. 
We do not use HDFS, it is an distributed filesystem intended replicated storage of very large (append-only) files -- many GBs, TBs, this makes it inappropriate for the files used in this study, which are at most 10Mb. 



\subsection{HarmonicIO}
We chose HarmonicIO for its simple P2P architecture. Its APIs are a simple abstraction around TCP sockets (in contrast to Apache Spark). Its container-based architecture provides a convenient way for scientists to encapsulate complex (and often fragile) software with a variety of dependent libraries, models and datasets. Docker containers are a useful `unit' of scientific computing code.


\section{EXPERIMENTAL SETUP \& BENCHMARKING TOOLS}\label{expsetup}

\begin{figure*}[h]
\includegraphics[width=\textwidth]{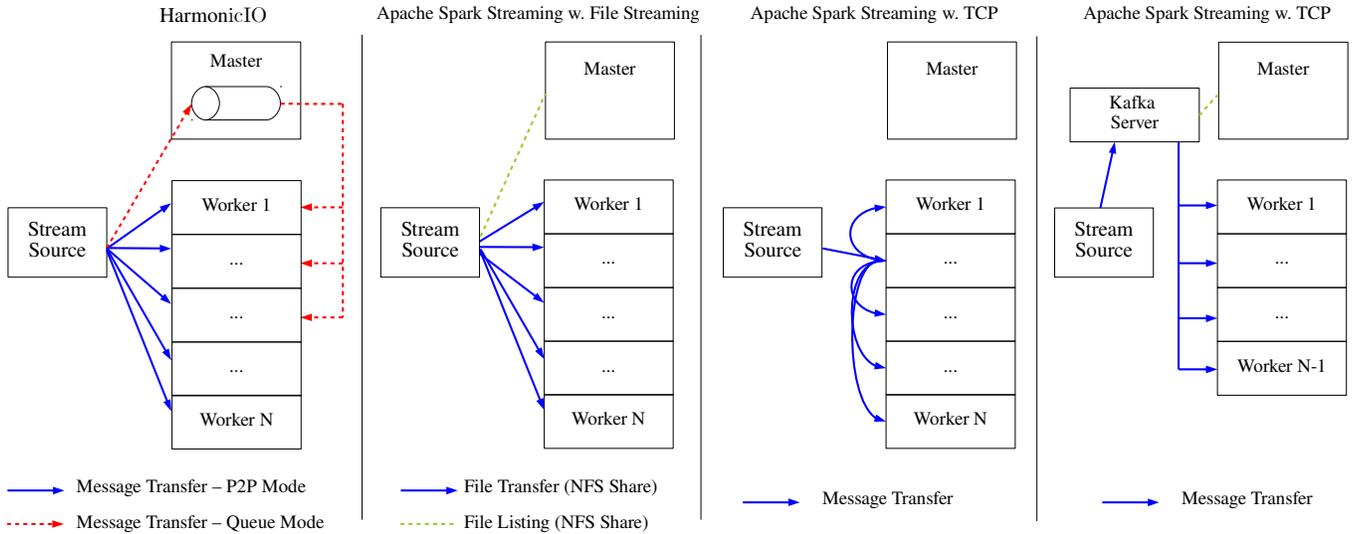}
\caption{Architecture / Network Topology Comparison between the frameworks and streaming source intregrations -- focusing on major network traffic flows. Note that with Spark and a TCP streaming source, one of the worker nodes is selected as `receiver' and manages the TCP connection to the streaming source. 
Note that neither the monitoring and throttling tool (which communicates with all components) are shown, nor is additional metadata traffic between nodes (for scheduling, etc.). Note the adaptive queue- and P2P-based message processing in HarmonicIO.}
\label{fig:arch-comp}
\end{figure*}

To explore the (message size, CPU cost per message) parameter space we developed benchmarking applications to run on Spark and HarmonicIO, able to process synthetic messages, and generate synthetic CPU load. These tools are publicly available at: \texttt{https://github.com/HASTE-project/benchmarking-tools}.
Fig.~\ref{fig:arch-comp} shows the various pipelines showing of HarmonicIO, and Apache Spark Streaming with file, Kakfa and TCP based integrations. The arrows show the busy network communication.
For each setup, 6 stream Processing VMs were used, each with 8 VCPUs and 16GB RAM (1 master, 5 workers). For the streaming source VM we used a 1 VCPU, 2GB RAM instance.  
These resources are similar to the experimental setup of \cite{xinApacheSparkFastest2014}, where 40 cores were used.
The maximum network bandwidth monitored using \texttt{iperf} was 1.4Gbit/s. 
Below, we describe the details of the experimental setup for each framework, and the approach used for determining the maximum frequency throughput.

\subsection{Apache Spark}
We created a \emph{streaming source} application, supporting TCP, Kafka, and file-based streaming integrations for ASS. The synthetic messages contain metadata describing the CPU load, so that both parameters (CPU load and message size) can be tuned in real time via the streaming source application.
To determine the maximum throughput, we adopt the approach of gradually increasing the message frequency (for a fixed (message\_size, CPU cost) pair) until a bottleneck is detected somewhere in the pipeline, then a binary search to find the maximum. A monitoring and throttling tool was developed for this purpose. Listing~1 shows a simplified view of the algorithm used to determine the maximum. To achieve this, it monitors and controls our streaming source application and the spark application, through a combination of REST APIs and log file analysis. 





\begin{figure}
\centering
\begin{scriptsize}
\begin{verbatim}
def find_max_f(msize, cpu_cost):
    max_known_ok_f <- 0
    min_known_not_ok_f <- null
    f <- f_last_run or default(msize, cpu_cost)
    while true:
        metrics = [spark_metrics(), strm_src_metrics()]
        switch heuristics(metrics):
        case sustained_load_ok: 
            f <- throttle_up(metrics, f)
        case too_much_load:
            f <- throttle_down(f)
        case wait_and_see:
            pass
     sleep

def throttle_up(metrics, f):
    max_known_ok_f <- f
    if min_known_not_ok_f == null:
        load = estimate_fraction_max_load(metrics)
        if load < 0.01: new_f <- f * 10
        elif load < 0.1: new_f <- f * 5
        elif load < 0.5: new_f <- int(f * 1.10)
        elif load < 0.8: new_f <- int(f * 1.05)
        else: new_f <- int(f * 1.05)
        if f == new_f:
            new_f <- f + 1
        return new_f
    else:
        return find_midpoint_or_done()

def throttle_down(f):
    min_known_not_ok_f <- f
    return find_midpoint_or_done()

def find_midpoint_or_done():
    if max_known_ok_f + 1 >= min_known_not_ok_f: 
        done(max_known_ok_f);
    else: 
        return int(mean(max_known_ok_f, 
                        min_known_not_ok_f))

\end{verbatim}
\end{scriptsize}
\caption*{Listing 1: Monitoring and Throttling Algorithm (Simplified). Until an interger upper bound is reached, increase piecewise linearly, then binary search.}
\end{figure}



This process is repeated for (message\_size, CPU cost) in a parameter sweep. 
We used a batch interval of 5 seconds, and a micro-batch interval of 150mS. Experimenting with other values had little impact on throughput. For the Spark File Streaming investigation, files are shared on the streaming server with NFS. 
Maximum throughput is reached when a bottleneck occurs somewhere in the system, as detected by the throttling tool:
\begin{itemize}
\item ASS taking too long to process messages.
\item There is a network bottleneck at the stream source.
\item For file streaming, ASS is taking too long to perform a directory listing.
\end{itemize}




\subsection{HarmonicIO}

In HarmonicIO, the maximum throughput is determined by measuring the time to stream and process a predefined number of messages for the given parameters.
We created a separate benchmarking application for HarmonicIO, which reads messages in our format. As with Spark, metadata describing the amount of CPU load is embedded in each message.


\section{RESULTS}\label{results}

\begin{figure}[h]
\begin{center}
\includegraphics[width=8cm]{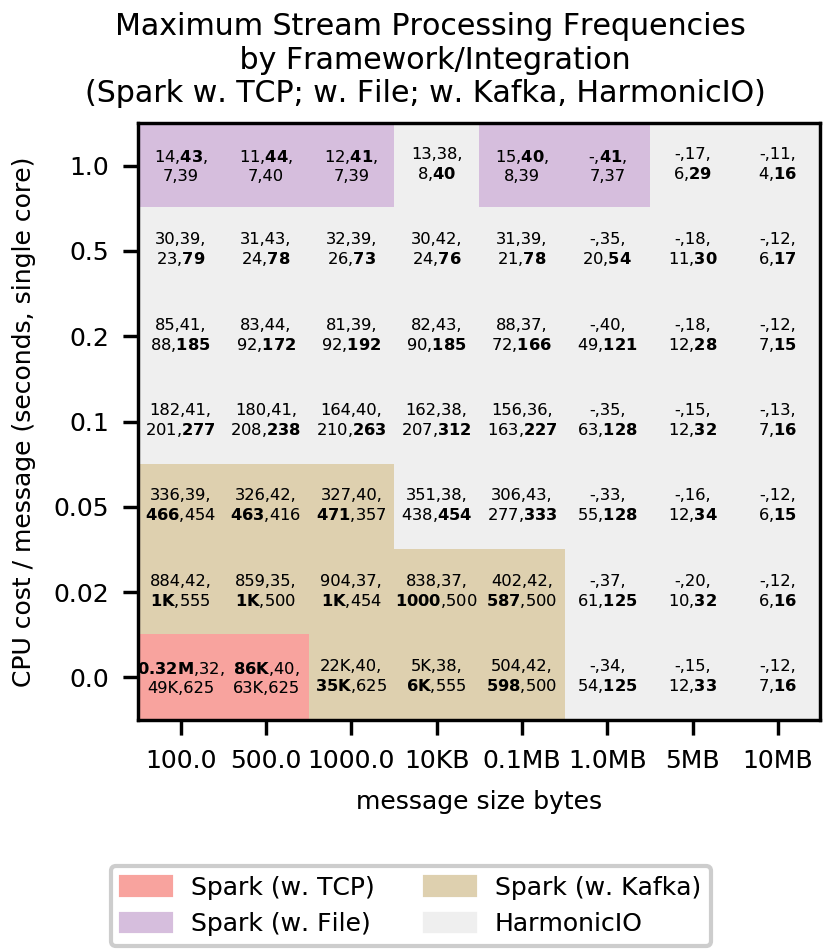}
\end{center}
\caption{Performance of Apache Spark (under TCP, File streaming, and Kafka integrations), and HarmonicIO, by maximum message throughput frequency over the domain under study. Under each setting, the highest frequency is shown in bold, with color-coding according to which framework/integration was able to achieve the best frequency. Compare with Fig.~\ref{fig:prediction}.}
\label{fig:blobs}
\end{figure}

\begin{figure*}[h]

\begin{center}
\includegraphics[width=0.9\textwidth]{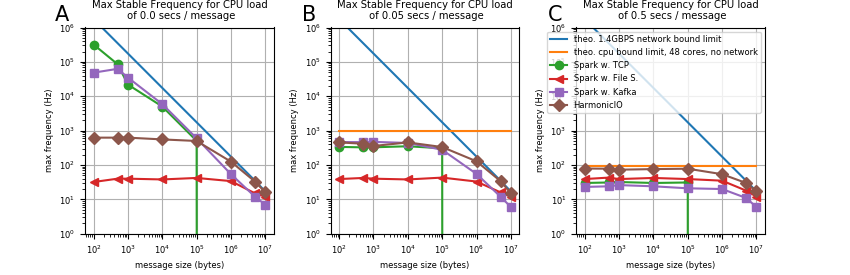}
\end{center}
\caption{Maximum stream processing frequencies for Spark (with TCP, File Streaming, Kafka) and HarmonicIO by message size; for selection of CPU cost/message.}
\label{fig:all-results}
\end{figure*}

\begin{figure}[h]
\begin{center}
\includegraphics[width=7cm]{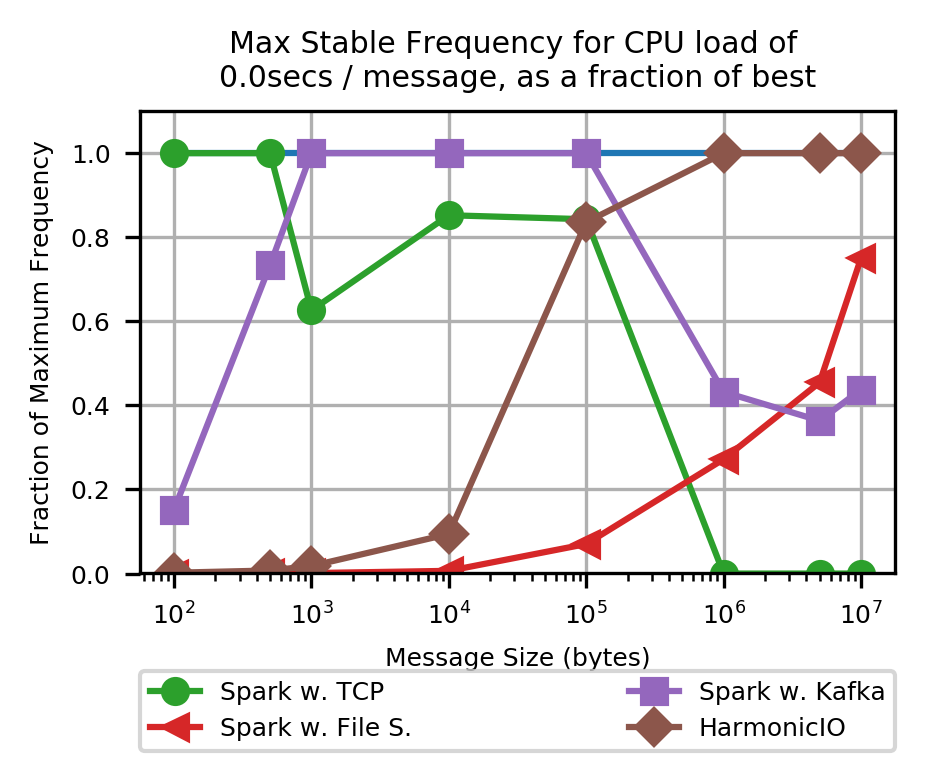}
\end{center}
\caption{Maximum frequency by message size, for Spark (with TCP, File Streaming, Kafka) and HarmonicIO under varying message size; normalized as a fraction of the best performing framework for the particular parameter values.}
\label{fig:result-normed}
\end{figure}

The maximum frequencies achieved by each framework (and stream integration setup), according to message size and per-message CPU load, are show in Fig.~\ref{fig:blobs}, color-coded according to the best performing framework. 
A subset of these results is presented again in Fig.~\ref{fig:all-results} and Fig.~\ref{fig:result-normed}, where results for particular CPU loads are shown in relation to CPU and network-theoretical bounds.

We summarize the results for each setup, before further discussion in relation to the intended use cases, architectures, and relation to theoretical bounds:

\textbf{Apache Spark Streaming with TCP}: 
This integration achieves very high frequency when message size and CPU load are small, consistent with previous studies. For 100 byte messages without CPU load; the pipeline was able to process messages at frequencies approaching 320KHz, meaning around 1.6M messages were processed by Spark in a 5 second batch. This can seen in the extreme lower-left of Fig.~\ref{fig:blobs}, and the results shown in Fig.~\ref{fig:all-results}.A. 
But performance degraded rapidly for larger message sizes, and under our benchmarks, it couldn't reliably handle messages larger than $10^5$ bytes at any frequency.

\textbf{Apache Spark Streaming with Kafka}:
This streaming pipeline performed well for messages less than 1MB, and CPU loads  less than 0.1 second/message, at the bottom-left of Fig.~\ref{fig:blobs}. Away from this region, performance degrades relative to other setups in this study, falling away from theoretical limits.

\textbf{Apache Spark Streaming with File Streaming}: 
This integration performed efficiently at low frequencies -- in regions tightly constrained by network and CPU-theoretic bounds (the top and right of Figs.~\ref{fig:prediction} and \ref{fig:blobs} and, and the results for higher message sizes in Fig.~\ref{fig:all-results}).

\textbf{HarmonicIO}:
Fig.~\ref{fig:blobs} shows that HIO was the best performing framework for the broad intermediate region of our study domain, 
for medium-sized messages (larger than 1.0MB), and/or CPU loads higher than 0.05 seconds/message. It matched the performance of file-based streaming with Apache Spark for larger messages and higher CPU loads.

\section{DISCUSSION}\label{discussion}

\subsection{Performance: Message Size \& CPU Load}



Maximum message throughput is bound by network and CPU bounds, which are inversely proportional to message size and CPU cost size respectively (assuming constant network speed, and number of CPU cores; respectively). The relative performance of different frameworks (and stream integrations) depend on these same parameters. Fig.~\ref{fig:blobs} shows that all frameworks also have specific, well-defined regions where they each perform the best. We discuss these regions in turn, moving outwards from the origin of Fig.~\ref{fig:blobs}.

Close to the origin, theoretical maximum message throughput is high, Fig.~\ref{fig:all-results}.A shows that for the smallest of messages, Spark with TCP streaming is able to outperform Kafka. 
For slightly larger message sizes (but less than 1MB), Kafka slightly outperforms Spark with TCP (it is better optimized for handling messages in this size range - its intended use case). For Kafka, Under the direct DStream integration with Spark, messages are transferred directly from the Kafka server to Spark workers. Yet, the Kafka server itself; having been deployed on its own machine, has theoretical network bounded throughput at half the network link speed (half the bandwidth for incoming messages, half for outgoing). Under TCP, we see a similar effect, with the spark worker nominated as receiver performing an analogous role. Consequently, Fig.~\ref{fig:all-results}.A shows that Spark with neither Kafka nor direct TCP can approach the theoretical network bound, as \emph{is} the case with other frameworks in some regions. This is consistent with the overview of the network topology shown in Fig.~\ref{fig:arch-comp}.


Moving further from the origin, into the region with CPU cost of 0.2-0.5 secs/message and/or medium size (1-10MB) -- HarmonicIO performs better than the Spark integrations under this study. 
It exhibits good performance -- transferring messages P2P makes means it is able to make very good use of bandwidth (when the messages are larger than 1MB or so); similarly, when there is considerable CPU load, its simplicity obviates spending CPU time on serialization, and passing messages multiple times among nodes. 
For large messages, and heavy CPU loads, this integration is able to approach closely to the network and CPU bounds -- its able to make cost-effective use of the hardware.

For the very largest messages, and the highest per-message CPU loads, the network and CPU bounds are very tight, and overall frequencies are very low (double digits). In these regions HarmonicIO performance is matched (or exceeded) by Spark with File Streaming. Both approaches are able to tightly approach the network and CPU theoretic bounds -- as shown in Fig.~\ref{fig:all-results}. Spark with file streaming performs slightly better in CPU-bound cases, with HarmonicIO performing slightly better in network-bound use cases (again, Fig.~\ref{fig:all-results}). 

\subsection{Performance, Frameworks \& Architecture}

This section draws together previous discussions, to summarize of the strengths and weaknesses of each framework and streaming source integration, their suitability for different use cases, with an emphasis on internals, and the network topology/architecture in each case.

Spark with TCP is able to process messages at very high frequencies in a narrow region; yet this performance is highly sensitive to message size and CPU load. Forwarding messages between workers yields allows message replication (and hence resilience), at a cost overall message throughput.
With heavy CPU loads, we see reduced performance of Spark with TCP relative to the other integrations with fewer cores available. 


Kafka is the best performing framework for slightly larger messages and CPU loads of less than ~0.05 secs/message.
The Direct DStream integration with Spark means that messages are being transferred directly from the Kafka server to the Spark workers (see Fig.~\ref{fig:arch-comp}).  
However, for larger messages (1-10MB), Kafka performs poorly (
Kafka is not intended for handling these file sizes). When the CPU load is more considerable, for our small cluster (48 cores), the overheads of Kafka (and Spark) reduce the overall CPU core utility -- there are simply less cores available for message processing. 
In these cases, performance is met or exceeded by HarmonicIO -- messages are transferred directly from the streaming source, so more cores are available for processing message content.


The results for Spark with File Streaming are quite different. The implementation polls the filesystem for new files, which no doubt works well for quasi-batch jobs at low polling frequencies -- order of minutes, with a small number of large files (order GBs, TBs, intended use cases for HDFS). However, for large numbers of much smaller files (MBs, KBs), this mechanism performs poorly. For these smaller messages, network-bound throughput corresponds to message frequencies of, say, several KHz (see Fig.~\ref{fig:all-results}). There frequencies are outside the intended use case of this integration, and the filesystem polling-based implementation is cumbersome for low-latency applications. 
The \texttt{FileInputDStream} is not intended to handle the deletion of files during streaming~\footnote{\texttt{https://issues.apache.org/jira/browse/SPARK-20568}}.

However, 
NFS is a simple integration, and allows message content to be transferred directly to where its processed on the workers, see Fig.~\ref{fig:arch-comp}. Consequently, where the theoretic bound on message frequency is down to double-digit frequencies, 
the integration is very efficient, and is the best performing framework at very high CPU loads (the top of Fig.~\ref{fig:blobs}). Each worker directly fetches the files it needs from the streaming source machine, making good use of network bandwidth, and robustly handles messages of arbitrary size. 


HarmonicIO is the best performing framework (in terms of maximum frequency) over much of the parameter space in this study, as shown in Fig.~\ref{fig:blobs}. 
It's peer-to-peer network topology (see Fig.~\ref{fig:arch-comp}) explains its ability to make good use of the network infrastructure, it has excellent performance in the network bound case, and its simplicity explains how this generalizes well over the rest of the domain. In scenarios with message sizes in 1-10MB range, or computationally intensive message analysis on medium clusters -- i.e. scientific computing applications, including low-latency microscopy image analysis -- overall throughput can be better using HarmonicIO. Plots in Fig.~\ref{fig:all-results} show that HarmonicIO is able to make excellent use of available network bandwidth for larger message sizes especially.
HarmonicIO achieved a maximum message transfer rate of ~625Hz, for the smallest, lightest messages (see Fig.~\ref{fig:blobs}) -- meaning in this domain it was easily beaten by Kafka and Spark with TCP stream integration. Fig.~\ref{fig:all-results}.A clearly shows this frequency bound -- making it unsuitable for enterprise use cases with high message frequencies.
HarmonicIO's dynamic P2P architecture switching between P2P and queuing modes is a simple approach to making effective use of network bandwidth: if all workers are busy, messages are sent to the queue, keeping the link to the streaming source fully utilized.

In summary, for very small messages (and lightweight processing), Spark with Kafka performs well (as does Spark with TCP streaming), and similarly for quasi-batch processing of very large files (at large polling intervals), Spark with file streaming integration performs well. These are the two typical enterprise stream analytics contexts: low-latency, high-frequency processing of small messages, and high-latency quasi-batch file-based processing of large message batches, log files, archives, etc.
This leaves a middle region -- messages 1-10 MB, and CPU intensive processing of small messages ($>$ 0.1 sec/message): conditions typical of many scientific computing use cases, including our motivating microscopy image processing use case. In this region, HarmonicIO is able to outperform the Spark streaming approaches benchmarked in this study. 
Our results show how replication and message forwarding have a cost in terms of processing (cores are used for forwarding messages), and network bandwidth. Benchmarking over a spectrum of processing loads makes this visible empirically.




\subsection{Features, Challenges \& Recommendations}\label{challenges}

This section briefly discusses the experiences with each framework setup, 
implementation issues, and challenges of measuring performance. Whilst HarmonicIO offers container isolation, and a simple and extensible code base, it lacks functionality for replication and fault handling with no delivery guarantees -- messages can be lost in the case of node failure. The current implementation has no reduce functionality. The simplicity of HarmonicIO means it has only a handful of configuration options.
Conversely, ASS's rich featureset creates complexity, especially for configuration. 

\section{CONCLUSIONS}\label{conc}
This study has confirmed Spark's excellent performance, consistent with earlier studies, albeit for use cases with small message size, and low message processing cost -- and quasi-batch loads at low frequencies. But, we also find that these `islands' of excellent performance do not generalize across the wider domain of use cases we studied. In particular, it was difficult for Spark to achieve good performance in the 1MB-10MB message size range (typical of microscopy image analysis, for example). 
Our results have shown the importance of choosing the a stream source integration appropriate for the message size and required performance.

By contrast, HarmonicIO performs well in this region, with good hardware utilization at low frequencies -- whilst not matching Spark for maximum frequency for the small message/cheap map function use case. Its simplicity makes its performance less sensitive to configuration settings, and parameters of the application load. Its lack of replication (and fault tolerance) allows for better throughput in some scenarios, but makes it unsuitable for handling valuable data not replicated elsewhere. Where datasets are generated deterministically, for numerical simulations for example, this may be less of a concern.
\emph{Out of the box} HarmonicIO outperforms many typical (or easily adopted) stream integrations with Spark in the intermediate region -- atypical of enterprise use cases for which Spark (and common stream integrations) are intended, and that this performance is robust over a greater domain (of message size, processing cost). This is in contrast to highly-tuned Spark deployments, which are likely to give high performance in perhaps narrower domains. HarmonicIO is sometimes able to achieve better (and more robust) performance because of its simplicity and lack of extensive functionality, making it much more lightweight -- and able to achieve better hardware utilization in some scenarios.

This paper has quantified the use cases where each framework integration can perform at or near theoretical bounds, and where each may perform poorly. We've highlighted key features differences between the frameworks. We hope our findings prove useful 
to designers of atypical stream processing applications, particularly within scientific computing.

\section{FUTURE WORK}\label{futureWork}



Our goal is to develop a SCaaS platform for 
scientific computing. The challenge is to combine the features of Spark and the robust performance of HarmonicIO into a single platform, or indeed investigate whether this is feasible. 


The challenge is to engineer a system which is able to adapt to users' performance, durability and economic cost requirements in in a smart and dynamic way, going beyond, say, merely tuning replication policies, so that instead the architecture itself is radically different depending on the use case. We could approach the problem `from above' and build meta-frameworks, and middleware,
or `from below' -- selecting integration with different underlying systems (and reconfiguring them) depending on the characteristics of the application load and the desired trade-offs. 
We feel this represents an open problem in cloud computing, and a key focus for our future research.


\section{Acknowledgements}
This work is funded by the Swedish Foundation for Strategic Research (SSF) under award no. BD15-0008, and the eSSENCE strategic collaboration on eScience. Computational resources were provided by the Swedish National Infrastructure for Computing via the SNIC Science Cloud (SSC)~\cite{ssc}, an OpenStack-based community cloud for Swedish academia. 




\bibliographystyle{apalike}
{
\bibliography{My_Library}}

\end{document}